\documentclass[12pt]{article}
\usepackage{comment}
\usepackage{amsmath}
\usepackage{amssymb}
\usepackage{graphicx}
\usepackage{subcaption}
\usepackage{url}
\usepackage[sort&compress, numbers, merge]{natbib}
\usepackage{physics}

\usepackage{todonotes}

\usepackage[colorlinks=true, linkcolor=blue, citecolor=blue,
urlcolor=black]{hyperref}

\begin{document}

\def\ps{\mathbf{p}}
\def\PS{\mathbf{P}}

\baselineskip 0.6cm

\def\simgt{\mathrel{\lower2.5pt\vbox{\lineskip=0pt\baselineskip=0pt
           \hbox{$>$}\hbox{$\sim$}}}}
\def\simlt{\mathrel{\lower2.5pt\vbox{\lineskip=0pt\baselineskip=0pt
           \hbox{$<$}\hbox{$\sim$}}}}
\def\simprop{\mathrel{\lower3.0pt\vbox{\lineskip=1.0pt\baselineskip=0pt
             \hbox{$\propto$}\hbox{$\sim$}}}}
\def\tr{\mathop{\rm tr}}
\def\SU{\mathop{\rm SU}}

\begin{titlepage}

\begin{flushright}
IPMU19-0188 
\end{flushright}

\vskip 1.1cm

\begin{center}

{\Large \bf 
Cosmological Constraint on Dark Photon from $N_{\rm eff}$ 
}

\vskip 1.2cm
Masahiro Ibe$^{a,b}$, 
Shin Kobayashi$^{a}$, 
Yuhei Nakayama$^{a}$ and
Satoshi Shirai$^{b}$
\vskip 0.5cm

{\it

$^a$ {ICRR, The University of Tokyo, Kashiwa, Chiba 277-8582, Japan}

$^b$ {Kavli Institute for the Physics and Mathematics of the Universe
 (WPI), \\The University of Tokyo Institutes for Advanced Study, \\ The
 University of Tokyo, Kashiwa 277-8583, Japan}
}

\vskip 1.0cm

\abstract{
A new U(1) gauge symmetry is the simplest extension of the Standard Model and has various theoretical and phenomenological motivations. 
In this paper, we study the cosmological constraint on the MeV scale dark photon. After the neutrino decoupling era at $T = \order{1}$~MeV, the decay and annihilation of the dark photon heats up the electron and photon plasma and accordingly decreases the effective number of neutrino $N_{\mathrm{eff}}$ in the recombination era. We derive a conservative lower-limit of the dark photon mass around 8.5~MeV from the current Planck data
if the mixing between the dark photon and ordinary photon is larger than $\order{10^{-9}}$.
We also find that the future CMB stage-I\!V experiments can probe up to 17~MeV dark photon.
}

\end{center}
\end{titlepage}

\section{Introduction}
\label{sec:intro}
The dark photon which stems from 
a new U(1) gauge symmetry is one of the 
simplest extensions of  the 
Standard Model (SM).
By assuming no SM fields are charged under the 
new U(1) gauge symmetry, it couples to the 
SM sector through the kinetic mixing with 
the gauge boson of the U(1)$_{\rm Y}$ in the SM
at the renormalizable level~\cite{Holdom:1985ag}.

The dark photon has various cosmological advantages.
For instance, the U(1) symmetry can be the origin of the stability of the dark matter.
Moreover, it is discussed that the dark matter self-interaction via the gauge interaction can solve the small scale structure problems of the collision-less dark matter~\cite{Kaplinghat:2015aga,Tulin_2018,Chu:2018fzy,Chu:2019awd}.
The dark photon also provides a portal to transfer excessive entropy in the dark sector into the SM sector before the neutrino decoupling~\cite{Blennow:2012de,Ibe:2018juk}.
In light of these features, the dark photon is gathering more and more attention and several new experiments are proposed to probe the sub-GeV dark photon  (see Ref.~\cite{Bauer_2018} for summary).

In this paper, we study the effective number of neutrino 
degrees of freedom,
$N_\mathrm{eff}$,
in the presence of the dark photon,
which is constrained by the cosmic microwave background (CMB) observations. 
As the MeV dark photon does not couple to the neutrinos, it would heat up only the electron-photon plasma
if it decays or annihilates after the neutrino decoupling. 
Such late-time energy injection can reduce $N_{\mathrm{eff}}$.
In previous analyses, the $N_\mathrm{eff}$ constraint puts an upper limit on the dark photon lifetime of $\tau_{\gamma'}<\order{1}$\,sec. 
As we will see, however, the MeV dark photon produced from the photon thermal bath can reduce $N_{\mathrm{eff}}$, thus it is constrained even in the case $\tau_{\gamma'}<\order{1}$\,sec.%
\footnote{The $N_{\mathrm{eff}}$ constraints on the $L_\mu-L_\tau$ gauge boson have been studied in Refs.~\cite{Kamada:2015era,Kamada:2018zxi,Escudero:2019gzq}.
In the case of the $L_\mu-L_\tau$ gauge boson in
the MeV range, it increases $N_{\mathrm{eff}}$ as the gauge boson decays into the neutrinos. }

In deriving the constraint, we solve the  Boltzmann equation of the dark photon  coupling to the photon, the electron, and the neutrino systems.
There, we use the full Boltzmann
equation of the momentum distribution of the dark photon which includes the Pauli-blocking and the Bose-enhancement effects.
This treatment is particularly important to derive the constraints on the scenario with freeze-in dark photon. 
As we will see, the freeze-in dark photon is excluded for $\varepsilon \gtrsim 10^{-9.5}$ and $m_\gamma \lesssim 8.5$\,MeV
by the latest Planck constraint~\cite{Aghanim:2018eyx}.
We also find that the stage-I\!V CMB
experiment~\cite{Abazajian:2016yjj} 
is sensitive to the dark photon mass 
up to about $17$\,MeV.
The constraint on the freeze-in dark photon provides the conservative and initial condition independent constraints, which can be generically applicable as long as we assume that the dark photon exists.
We also discuss the constraint in 
the scenarios where 
the dark photon decouples from the SM thermal bath
in the early universe.

The organization of the paper is as follows.
In Sec.~\ref{sec:setup}, we summarize the relevant 
properties of the dark photon.
In Sec.~\ref{sec:analysis}, we provides the 
full Boltzmann equation of the momentum distribution of the dark photon.
In Sec.~\ref{sec:constratins}, we show the 
constraints on the dark photon in the freeze-in scenario as well as the scenarios with early decoupled dark photon.
The final section is devoted to  discussions.
\section{The model of dark photon}
\label{sec:setup}
The massive dark photon has the kinetic mixing interaction with the QED photon,
\begin{align}
    \mathcal{L}_{\mathrm{mix}} = -\frac{1}{4}F^{\mu\nu}F_{\mu\nu} - \frac{1}{4}F'^{\mu\nu}F'_{\mu\nu}-
    \frac{\varepsilon}{2}F^{\mu\nu}F'_{\mu\nu} + \frac{1}{2} m^2_{\gamma'}A'^\mu A'_\mu  + e A_\mu J^\mu_{\mathrm{QED}} \ . 
    \label{eq:interaction}
\end{align}
Here, $F_{\mu\nu}$ ($F'_{\mu\nu}$) represents the field strength of the QED (dark) photon, $A^\mu\, (A'^\mu)$ is the SM (dark) photon field,
and $J_{\mathrm{QED}}$ is the QED current.
The gauge coupling constant of QED 
is given as $e$, 
while $m_{\gamma'}$ and $\varepsilon$ are the dark photon mass
and the mixing parameter, respectively.
Throughout this paper, we assume that the kinetic mixing is tiny, $\varepsilon \ll 1$.%
\footnote{The tiny kinetic mixing 
term can be naturally obtained 
when the U(1) gauge symmetry to which the dark photon associates is embedded into a non-Abelian gauge group at a high energy (see e.g, Refs.~\cite{Ibe:2018tex,Ibe:2019ena}).}
The redefinition of the QED photon field eliminates the kinetic mixing term, which induces the dark photon interaction, $\varepsilon e A_\mu'J^\mu_{\mathrm{QED}}$. 
Accordingly, the partial decay rate of the dark photon into a 
pair of the electron and positron is given by
\begin{align}
    \Gamma_{\gamma'\to e^+e^-} = \frac{1}{3}\alpha \varepsilon^2 m_{\gamma'}\left(1+\frac{2m_e^2}{m_{\gamma'}^2}\right)\sqrt{1 - \frac{4m_e^2}{m_{\gamma'}^2}}\ , 
\end{align}
where $\alpha = e^2/4\pi$ is the QED fine structure constant and $m_e = 0.511\, \mathrm{MeV}$ is the electron mass.
Since the dark photon coupling to the neutrinos are suppressed,
it heats up only the electron-photon plasma
if it decays or annihilates after the neutrino decoupling.
This effect reduces $N_{\mathrm{eff}}$, which can be
constrained by the CMB observations.

Let us discuss the production of the dark photon in the early 
Universe.
There are various production mechanisms of the dark photon depending on the cosmological history as well as 
the underlying dark sector to which 
the dark photon belongs.
For example, the dark sector may 
also contain a dark Higgs and dark matter.
In this work, we focus on two mechanisms: production from the SM thermal plasma (freeze-in mechanism) and that from the dark sector thermal bath (freeze-out mechanism).%
\footnote{In both scenarios, we assume that no dark sector particles such as
the dark Higgs appear 
in the MeV region except for the dark photon. 
The presence of additional light particles makes thermal history
more complicated. 
}
In the first case, the 
dark photon contribution is solely determined by $\varepsilon$ and $m_{\gamma'}$. In the second case,
it depends on other 
parameters in the dark sector.

\begin{figure}[t]
\centering
	\begin{minipage}{0.3\hsize}
	\centering
	\includegraphics[width=0.9\hsize,clip]{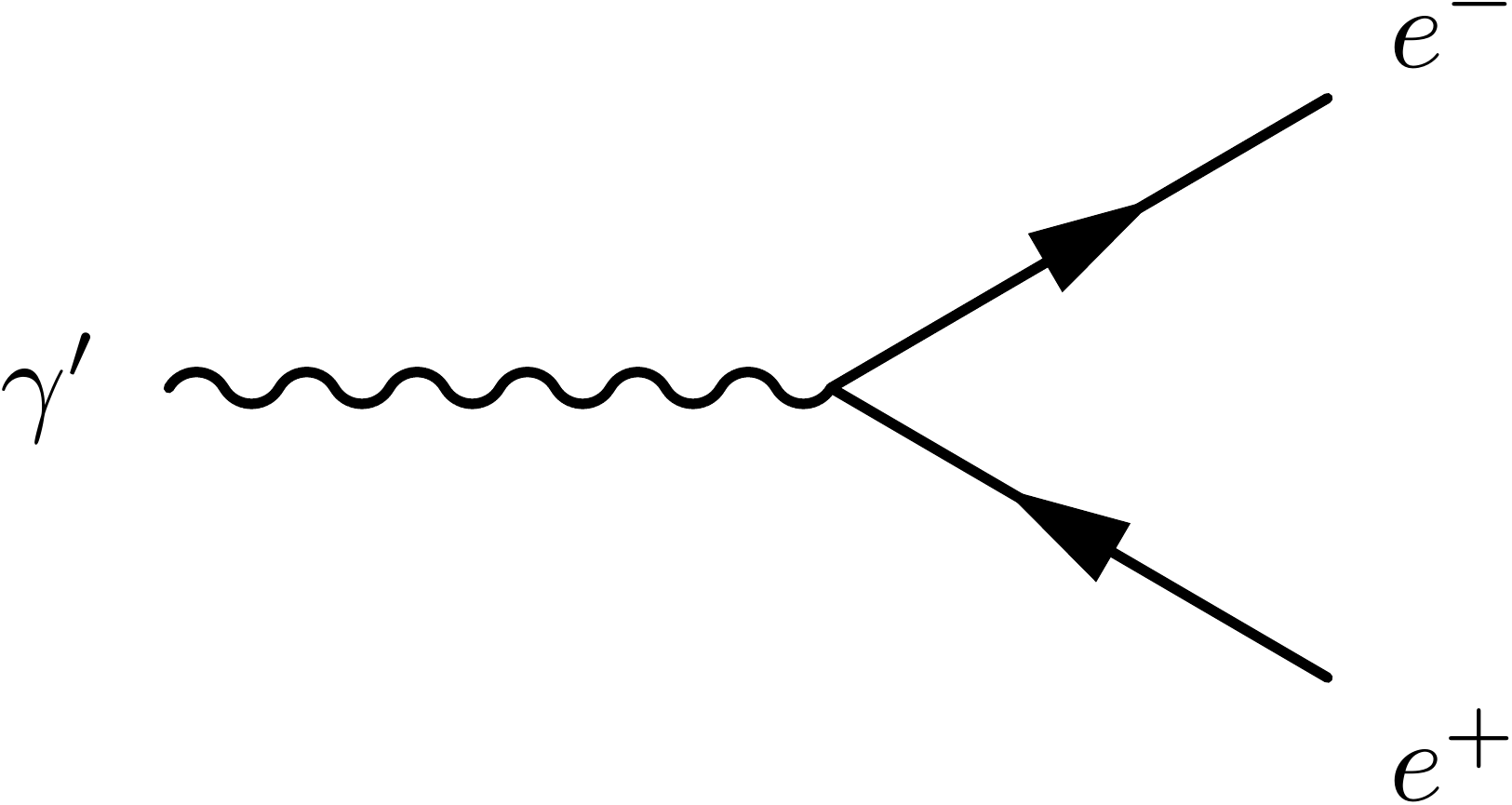}
    \end{minipage}
    \begin{minipage}{0.3\hsize}
	\centering
	\includegraphics[width=0.9\hsize,clip]{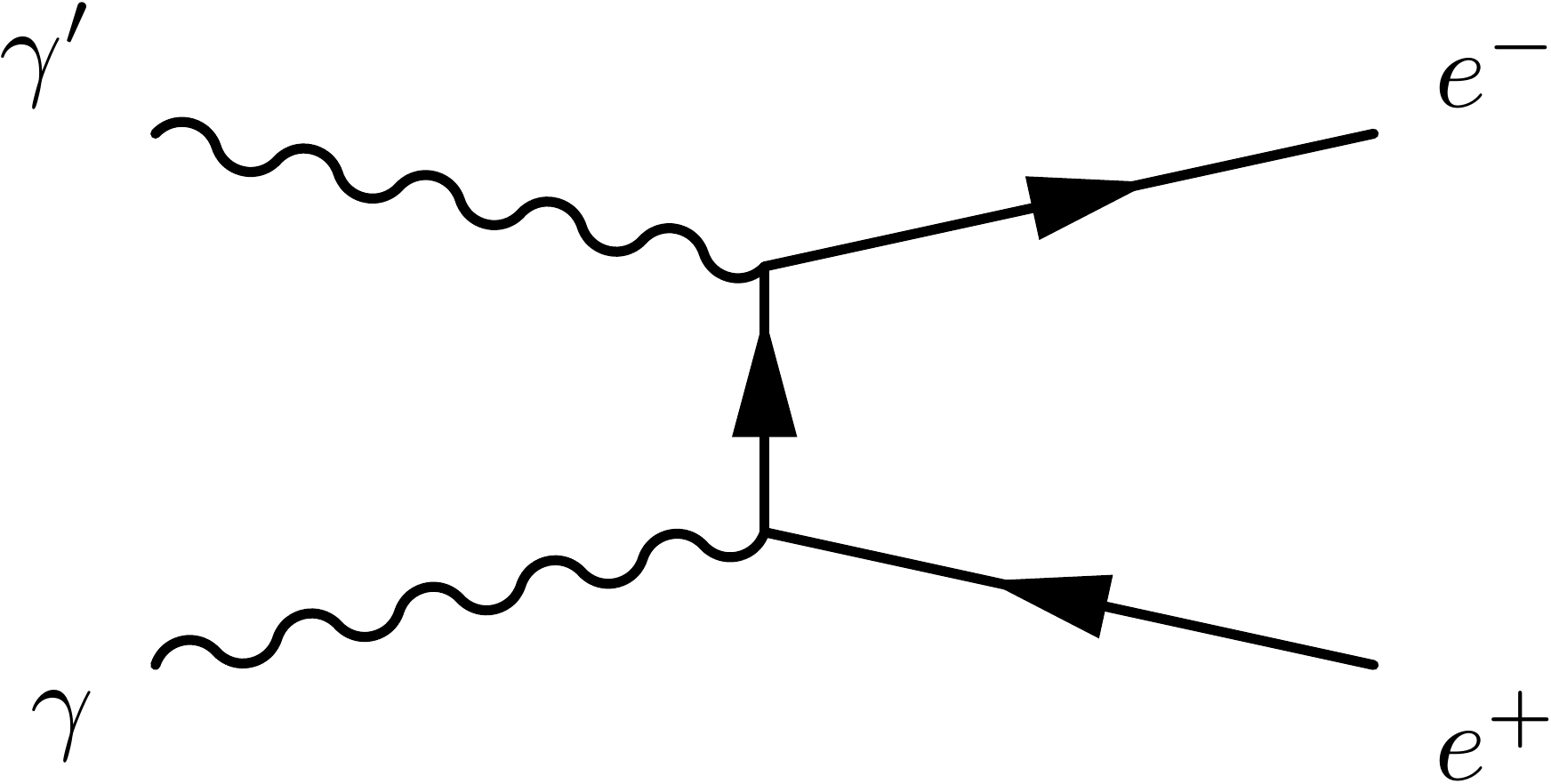}
    \end{minipage}
    \begin{minipage}{0.3\hsize}
	\centering
	\includegraphics[width=0.9\hsize,clip]{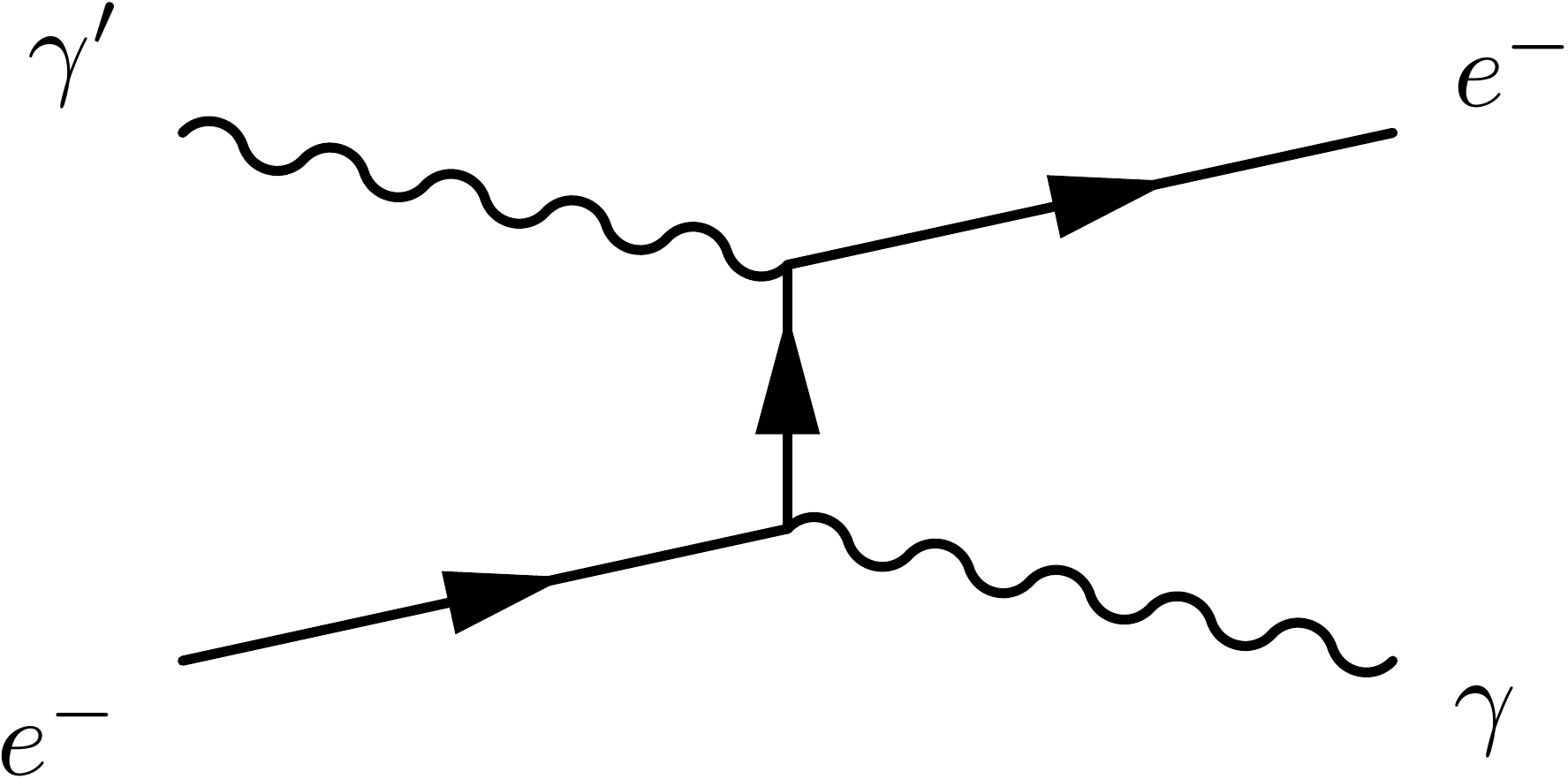}
    \end{minipage}
	\caption{
	The Feynman diagrams relevant for the dark photon decay (left), annihilation (middle) and the Compton-scattering like process (right).}
	\label{fig:diagrams}
\end{figure}

The first one is the thermal freeze-in mechanism, in which the
dark photons are produced from the SM plasma via the interaction  in Eq.\,\eqref{eq:interaction}.
For example, the (inverse) 
decay process $e^++e^- \leftrightarrow \gamma'$ and the scattering processes such as $e^++e^- \leftrightarrow \gamma+\gamma'$ contribute to the freeze-in production (see Fig.\,\ref{fig:diagrams}).

The thermal averaged production rates are 
$\langle \Gamma_{e^++e^-\to\gamma'} \rangle
\sim \alpha\varepsilon^2 m_{\gamma'}^2/T$
and $\langle \Gamma_{\mathrm{scattering}} \rangle \sim \alpha^2\varepsilon^2T$, respectively,
where $T$ is the temperature of the SM thermal bath.
Thus, the inverse-decay and the scattering productions become 
larger than the Hubble 
expansion rate $H$ only when the $T$ becomes lower than $T_{\mathrm{ID}} \simeq (\alpha\varepsilon^2m_{\gamma'}^2 M_P)^{1/3}$
and $T_{\mathrm{SC}}\simeq \alpha^2\varepsilon^2 M_P$, respectively.
Here, $M_P\simeq 2.4\times 10^{18}$\,GeV is the reduced Planck scale.

Thus, the dark photon is not in thermal equilibrium with the SM thermal bath 
through the kinetic mixing 
until $T$ becomes lower than  
$T_{\mathrm{FI}}= \max[T_{\mathrm{ID}},T_{\mathrm{SC}}]$.

In this scenario, the produced dark photons 
reaches to the thermal equilibrium with the electron-photon plasma for 
$T_{\mathrm{FI}} \gg \max[m_e,m_{\gamma'}]$.
It should be emphasized that this production mechanism scarcely depends on the reheating temperature $T_R$ of the Universe after the primordial inflation as far as $T_R > m_e$ and $m_{\gamma'}$.
This is because the dark photon production from the electron-photon plasma is dominated at  $T \sim \max\left[m_e,m_{\gamma'}\right]$.
This mechanism guarantees the minimum amount of the dark photon in the early Universe, regardless of the initial condition of the Universe and details of the dark sector.
Therefore we can obtain the most conservative constraint if we consider only the freeze-in contribution.
Hereafter, we refer this conservative case as the ``freeze-in scenario."

The second one is 
the freeze out mechanism of the dark sector,
where the dark sector used to be thermalized with the SM sector through interactions other than
the kinetic mixing in the very early Universe.
For instance, if the dark sector contains heavy dark Higgs particles, $H_D$, which have couplings to the SM Higgs, $H_{\mathrm{SM}}$, i.e. $|H_D|^2 |H_{\mathrm{SM}}|^2$,
the dark sector including the dark photon can be thermalized and have the same temperature as the SM sector.
As the temperature gets lower, the dark sector 
decouples from the SM sector at $T_D$.
The dark sector decoupling temperature is 
independent of $\varepsilon$, and we take it as 
a free parameter.

If we assume the case of sudden dark sector decoupling and if the Hubble rate at that time is much greater than $\Gamma_{\gamma'}$, the momentum distribution of the dark photon in$ \max\left[m_{e},m_{\gamma'},T_{\mathrm{FI}}\right] \ll  T < T_D$ is given by,
\begin{align}
\label{eq:decoupled}
f_{\mathrm{\gamma'}}(p_{\gamma'}) = \frac{1}{e^{\sqrt{m_{\gamma'}^2 + (p^{D}_{\gamma'})^2}/T_{D}}-1} \text{ with }
p^D_{\gamma'} = \left(\frac{g_{*S}(T_D)T_D^3}{g_{*S}(T_{\gamma e})T^3}\right)^{1/3} p_{\gamma'},
\end{align}
where  $g_{*S}$ is given by the entropy density of the SM thermal of temperature $T$: $s =2\pi^2 g_{*S}(T)T^3/45$.
Due to the large effective massless degrees of freedom of the SM, the number density of the dark photon is diluted by $g_{*S}(T)/g_{*S}(T_D)$ compared to the thermalized case.
We refer this case as the ``dark sector freeze-out scenario", where the initial dark photon distribution is determined by the dark sector decoupling temperature $T_D$.

Before closing this section, let us comment on the 
dark photon decay for $m_{\gamma'} < 2 m_e$.
In this regime, the main mode of the dark photon decay
is either the one into the three photons or the 
one into the neutrinos through the mixing with the SM $Z$-boson.
The decay rate into the three-photon
is given by,
\begin{align}
    \Gamma_{\gamma'\to 3\gamma} =
\frac{17\varepsilon^2 \alpha^4}{2^73^65^3\pi^3}\frac{m_{\gamma'}^9}{m_e^8}\mathcal{F}(m_{\gamma'}^2/m_e^2)\ ,
\label{eq:3gamma}
\end{align}
where the prefactor corresponds to the decay rate in the Euler-Heisenberg limit~\cite{Pospelov_2008},
while the enhancement factor $\mathcal{F}(x)$ is given in Ref.~\cite{McDermott_2018}.%
\footnote{Here, the enhancement factor is normalized so that $\lim_{x\to 0}\mathcal{F}(x)=1$.}
The decay into the neutrino is induced by the kinetic mixing to the SM $Z$-boson,
\begin{align}
\mathcal{L} = \frac{\varepsilon\tan\theta_W}{2}Z^{\mu\nu}F_{\mu\nu}'\ ,
\end{align}
where $Z^{\mu\nu}$ denotes 
the field strength of the SM $Z$-boson, and $\theta_W$ is the Weinberg angle.
After eliminating the kinetic mixing term and diagonalizing the mass term,
the dark photon has the coupling to the SM neutral current, 
\begin{align}
\mathcal{L} = \varepsilon g \frac{\sin\theta_W}{\cos^2\theta_W}
\frac{m_{\gamma'}^2}{m_{Z}^2} A_\mu'J_Z^\mu \ ,
\end{align}
where $J_Z^\mu$ is the neutral current in the SM.
Accordingly, the dark photon decay rate into a pair of the neutrinos is given by,
\begin{align}
\Gamma_{\gamma'\to2\nu} = \frac{\varepsilon^2 g^2\tan^2\theta_W }{96\pi \cos^2\theta_W} \frac{m_{\gamma'}^5}{m_Z^4}\ ,
\end{align}
where $m_Z$ is the mass of the SM $Z$-boson and $g$ is the gauge coupling constant of ${\rm SU}(2)_{\rm L}$ gauge interaction of the SM.
Thus, we find that the three photon mode is dominant for $m_{\gamma'}> \order{10}$\,keV, while 
the neutrino mode is dominant for a lighter dark photon.

\section{Boltzmann Equations}
\label{sec:analysis}

In this section, we summarize the Boltzmann equations 
relevant for the calculation of $N_{\mathrm{eff}}$.
The equation for the momentum distribution of the dark photon, $f_{\gamma'}(p_{\gamma'})$, is written as
\begin{align}
    \pdv{f_{\gamma'}}{t} - H p \pdv{f_{\gamma'}}{p} & = - G_{ \gamma' \leftrightarrow e}(p, T_{\gamma e})\ ,  \label{eq:boltzmann}\\
    G_{ \gamma' \leftrightarrow e}(p_{\gamma'},T) &=\frac{m_{\gamma'} \Gamma_{\gamma'} (1 + \varphi(T_{\gamma e},p_{\gamma'})) }{E_{\gamma'}} \left(f_{\gamma'} - f_{\gamma'}^{\rm eq} (p_{\gamma'},T_{\gamma e})\right)\ .
    \label{eq:DecayInverseDecay}
\end{align}
Here, $f_{\gamma'}^{\rm eq} (p_{\gamma'},T)$ is the Bose-Einstein (BE) 
distribution: $f_{\gamma'}^{\rm eq} (p_{\gamma'},T)= 1/(\exp(E_{\gamma'}/T) - 1)$.
In the following, we use $T_{\gamma e}$ to 
specify the temperature of the electron-photon
thermal plasma.
The function $G_{\gamma' \leftrightarrow e}$ represents the collision term for the decay of $\gamma'$ and its inverse process.
In deriving the collision term of $f_{\gamma'}$, we use 
the BE distribution for the photon distribution, and 
the Fermi-Dirac (FD) distribution for the electron and the positron distributions.
In this case, the function $\varphi$ is given by
\begin{align}
\label{eq:phi}
    \varphi(T,p_{\gamma'}) = \frac{m_{\gamma'} T}{p_{\gamma'} p^0_e}
    \log\left( 
    \frac{e^{E_{\gamma'} E^0_e/(T m_{\gamma'})} + e^{ -p_{\gamma'} p^0_e/(T m_{\gamma'}) }}{
    e^{E_{\gamma'} E^0_e/(T m_{\gamma'})} + e^{ p_{\gamma'} p^0_e/(T m_{\gamma'}) }
        }
    \right)\ ,
\end{align}
where $p^0_e =  \sqrt{m_{\gamma'}^2 - 4 m_e^2}/2$ and  $E^0_e = m_{\gamma'}/2$ are the momentum and energy of the electron at the rest frame of the dark photon, $\gamma'$.
The derivation is shown in the Appendix~\ref{sec:DP decay}.

In addition to the decay and the inverse decay processes, we also take into account $\gamma' + \gamma \leftrightarrow e^- + e^+$ and $\gamma' + e^\pm \leftrightarrow \gamma + e^\pm$.
Such processes are subdominant for $m_{\gamma'} > 2 m_e$
as they are suppressed by an additional power of $\alpha$ compared with Eq.\,\eqref{eq:DecayInverseDecay}.
For $m_{\gamma'} < 2 m_e$, on the other hand,
they are the main production/annihilation processes of the dark photon, where the decay and the inverse decay 
are ineffective.
We show the Boltzmann equation for those processes in the Appendix~\ref{sec:scattering}.

When we calculate the collision terms of these processes, we encounter two types of the infrared (IR) divergences for $m_{\gamma'} > 2 m_e$
(see the Appendix~\ref{sec:scattering}).
One of which stems from the Bose enhancement of the scattered photon,
and the other is from the soft photon emission/absorption.%
\footnote{Due to the Bose enhancement, the tree-level contributions to the collision term have a linear IR divergence.}
In order to take care of these IR divergences appropriately, 
we have to add up 1-loop diagrams of the dark photon decay 
and the tree-level diagrams of the soft photon emission/absorption with finite temperature fermion propagators~\cite{Donoghue:1983qx}.
However, as we noted before, contributions from $\gamma' + \gamma \leftrightarrow e^- + e^+$ and $\gamma' + e^\pm \leftrightarrow \gamma + e^\pm$ are subdominant for $m_{\gamma'}>2m_e$ which
is the region of our main interest.
Therefore, in the calculation of the collision terms of these two processes, we simply introduce a thermal mass effects as a soft photon mass cut-off to avoid the IR divergences.
The detail of the prescription is shown in Appendix~\ref{sec:scattering}.
For $m_{\gamma'} < 2 m_e$ where these processes are dominant, on the
other hand, we do not have IR singularities from the tree-level contributions.

To estimate $N_{\mathrm{eff}}$, we need to solve the Boltzmann equations
for the SM sector simultaneously.
In our analysis, we follow Ref.~\cite{Escudero:2018mvt},
which allows an efficient and precise estimation of $N_{\mathrm{eff}}$.
There, the Boltzmann equations for SM part are given by
\begin{align}
    \frac{d \rho_{\gamma e}}{dt}  &= -3H(\rho_{\gamma e} + p_{\gamma e} ) +   C_{\gamma' \leftrightarrow e}(T_{\gamma e}) + C_{e \leftrightarrow \nu_e }(T_{\gamma e}, T_{\nu_{e}}) + 2 C_{e \leftrightarrow \nu_{\nu,\tau} }(T_{\gamma e}, T_{\nu_{\mu,\nu}})\  , \\
    C_{\gamma' \leftrightarrow e}(T_{\gamma e}) &= \frac{3}{(2\pi)^3}\int d^3 p_{\gamma'} E_{\gamma'}  G_{ \gamma' \leftrightarrow e}(p_{\gamma'},T_{\gamma e})\ , \\
    \frac{d \rho_{\nu_e}}{dt} & = -4H \rho_{\nu_e} -   C_{e \leftrightarrow \nu_e}(T_{\gamma e}, T_{\nu_{e}}) + 2 C_{\nu_e \leftrightarrow \nu_{\mu, \tau} }(T_{\nu_e}, T_{\nu_\mu})\ ,\\
    \frac{d \rho_{\nu_{\mu,\tau}}}{dt} & = -4H \rho_{\nu_{\mu,\tau}}  - 2 C_{e \leftrightarrow \nu_{\nu,\tau} }(T_{\gamma e}, T_{\nu_{\mu,\nu}}) - 2C_{\nu_e \leftrightarrow \nu_{\mu, \tau} }(T_{\nu_e}, T_{\nu_\mu})\ .
\end{align}
Here, $\rho_{\gamma e}=\rho_\gamma + \rho_e + \delta \rho$ and $p_{\gamma e} = p_\gamma + p_e +\delta P$ each represents the electron-photon plasma density and pressure, and $\rho_{\nu_{\mu, \tau}}$ is the sum of the densities of $\nu_\mu$ and $\nu_\tau$.%
\footnote{Following the analysis in Ref.~\cite{Escudero:2018mvt},
we assume that the temperatures of $\nu_\mu$ and $\nu_\tau$
are equal with each other, which is justified as their oscillation 
rate is larger than the Hubble expansion rate for $T =  \order{1}$\,MeV.}

Thermodynamical quantities $\rho_i$ and $p_i$ are calculated from the relation
\begin{align}
    \rho_i &= \int \frac{g_i d^3p_i}{(2\pi)^3}E_i\frac{1}{e^{E_i/T_i}\pm 1}\ ,\\
    p_i &= \int \frac{g_i d^3p_i}{(2\pi)^3}\frac{p_i^2}{3E_i}\frac{1}{e^{E_i/T_i}\pm 1}\ ,
\end{align}
where $g_i$ is the degrees of freedom and the sign of denominator depends on the statistics of the particle.
$\delta \rho$ and $\delta P$ are the QED loop corrections to the energy density and the pressure of the electron-photon plasma calculated as~\cite{Escudero:2018mvt, Heckler:1994tv, Mangano:2001iu, Fornengo:1997wa}, 
\begin{align}
    \delta \rho &= - \delta P + T_{\gamma e}\frac{d}{d T_{\gamma e}}\delta P \ , \\
    \delta P &= -\int_{0}^{\infty}\frac{dp}{2\pi^2}\left[ \frac{p^2}{\sqrt{p^2 + m_e^2}} \frac{\delta m_e^2(T_{\gamma e})}{e^{\sqrt{p^2 + m_e^2}/T}+1} + \frac{p}{2}\frac{\delta m_\gamma^2(T_{\gamma e})}{e^{p/T_{\gamma e}}-1}\right]\ , \\
    \delta m_e^2(T) &= \frac{2\pi\alpha T^2}{3} + \frac{4\alpha}{\pi}\int_{m_e}^{\infty} dE \frac{\sqrt{E^2 - m_e^2}}{e^{E/T} + 1}\ ,\\
    \label{eq:thermal photon mass}
    \delta m_\gamma^2(T) &= \frac{8\alpha}{\pi}\int_{m_e}^{\infty} dE \frac{\sqrt{E^2 - m_e^2}}{e^{E/T} + 1}\ .
\end{align}
With these quantities, the Hubble expansion rate is defined as
\begin{align}
    H&=\frac{\sqrt{ \rho_{\gamma'} +  \rho_{\gamma e} +  \rho_{\nu_e} +  \rho_{\nu_{\mu}} + \rho_{\nu_{\tau}}  + \rho_{\rm heavy}(T_{\gamma e})  } }{\sqrt{3} M_{P}}\ ,\\
    \rho_{\gamma'} &= \frac{3}{(2\pi)^3}\int d^3 p_{\gamma'} E_{\gamma'}  f(p_{\gamma'})\ ,
\end{align}
where $\rho_{\rm heavy}(T_{\gamma e}) $ represents the energy density from heavier SM particles (e.g., muon and hadrons) other than the electron, neutrino and photon.
We assume they have the same temperature as the electron and photon sector and adopt the result of Ref.~\cite{Saikawa:2018rcs} for the numerical estimation.

The initial conditions 
of the Boltzmann equation are set at $T_{\mathrm{init}} = T_{\gamma e} = T_{\nu_{e,\mu,\tau}} = 300$\,MeV.
For the freeze-in scenario, we take $f_{\gamma'} = 0$. For the dark sector freeze-out scenario,
we take $f_{\gamma'}$ to be the one in Eq.\,\eqref{eq:decoupled}.
Note that for $T_{\mathrm{FI}} \gg T_{\mathrm{init}}$, the above initial conditions
are not proper since the dark photon has been
thermalized by the kinetic mixing interaction
at a temperature higher than $T_{\mathrm{init}}$.
However, even if we set the above (improper)
initial conditions, the dark photon
distribution is immediately thermalized below $T_\mathrm{init}$.
In fact, the solution of the Boltzmann equation reaches to the
thermal distribution instantaneously.
Thus, the resultant $N_{\mathrm{eff}}$ constraint is independent of the initial condition for a large $\varepsilon$, i.e. $T_{\mathrm{FI}} \gg T_{\mathrm{init}}$.

After solving the Boltzmann equation, $N_{\mathrm{eff}}$ is given by
\begin{align}
    N_{\mathrm{eff}} = \frac{8}{7}\left(\frac{11}{4}\right)^{4/3}
\frac{\rho_{\nu_{e}}(T_{\nu_e})+\rho_{\nu_{\mu}}(T_{\nu_\mu}) + \rho_{\nu_{\mu}}(T_{\nu_\tau})}{\rho_{\gamma}(T_{e\gamma})}\ ,
\end{align}
which is evaluated at the temperature much below the dark photon decay 
temperature and the electron mass.
In our numerical analysis, we 
stop solving the Boltzmann equation at $T_{\gamma e}=0.5$\,keV below which the 
double Compton scattering becomes ineffective~\cite{Jeong:2014gna}.
Below this temperature, interactions with neutrinos are already decoupled, thus the further evolution does not affect the value of $N_{\mathrm{eff}}$, in the case dark photons decay above $0.5\, \mathrm{keV}$.
The dark photon which decays (or annihilates) 
below $T_{\gamma e}\lesssim 0.5$\,keV is constrained by 
the CMB spectrum distortion~\cite{Zeldovich:1969ff,Sunyaev:1970eu},
which has been applied to the dark photon in Refs.~\cite{Fradette:2014sza,Redondo:2008ec}.

Several comments are in order.
First, although we assume the BE or FD statistics for the photon 
and the electron/neutrinos in the calculation of thermodynamical quantities, 
we use the approximation that electrons and neutrinos obey the Maxwell-Boltzmann distribution and the Pauli blocking effects are negligible in the calculation
of $C_{e \leftrightarrow \nu_i}$ and $C_{\nu_i \leftrightarrow \nu_j}$.
We also ignore the masses of the electrons and the neutrinos in the calculations of the collision terms.
Those approximations are validated in~\cite{Escudero:2018mvt},
which affect 
$N_{\mathrm{eff}}$ less than $1$\%.
The explicit forms of these collision terms are given in the Appendix~\ref{sec:neutrino collision}.

Due to those approximations, the 
SM limit, i.e. $\left.N_{\mathrm{eff}}\right|_{\varepsilon= 0,f_{\gamma'} = 0}=3.053$ \cite{Escudero:2018mvt}, is slightly different from the state-of-the-art evaluation 
in the SM~\cite{deSalas:2016ztq}, 
$ N_{\mathrm{eff}}^{\mathrm{SM}} = 3.045$.
In the following, we define ${\mit \Delta}N_{\mathrm{eff}}$ by,
\begin{align}
    {\mit \Delta}N_{\mathrm{eff}} 
    = \left. N_{\mathrm{eff}} - N_{\mathrm{eff}}\right|_{\varepsilon= 0,f_{\gamma'} = 0} \ ,
\end{align}
and redefine $N_{\mathrm{eff}}$ by 
\begin{align}
   N_{\mathrm{eff}} = {\mit \Delta}N_{\mathrm{eff}} + N_{\mathrm{eff}}^{\mathrm{SM}}
\label{eq:shift}
\end{align}
where $ N_{\mathrm{eff}}^{\mathrm{SM}} = 3.045$ in the SM~\cite{deSalas:2016ztq}.

\begin{figure}[tbp]
	\centering
	\subcaptionbox{\label{fig:DP_evolution} $\rho_{\gamma'}$ }{\includegraphics[width=0.47\textwidth]{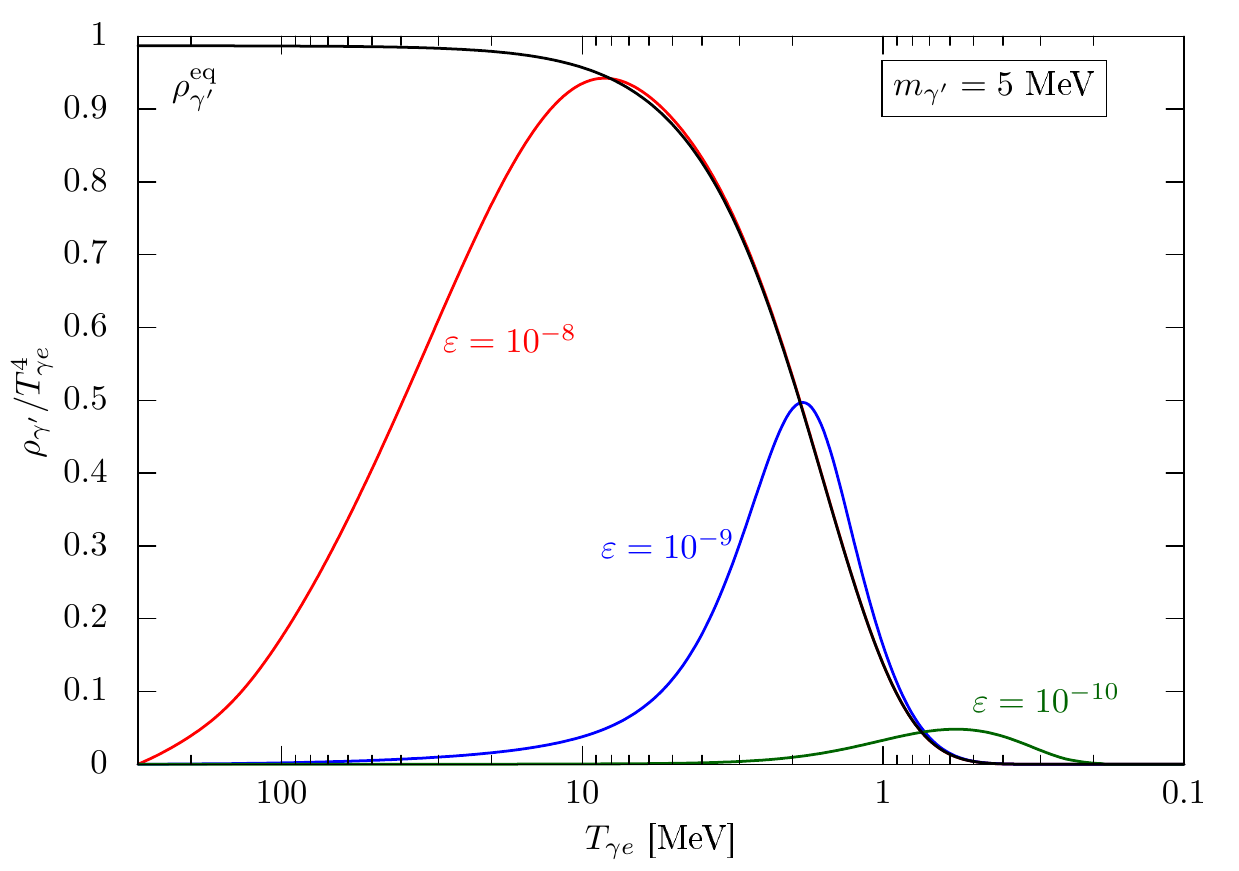}}
	\subcaptionbox{\label{fig:temperature_evolution}$T_{\gamma e}/T_{\nu_e}$}{\includegraphics[width=0.47\textwidth]{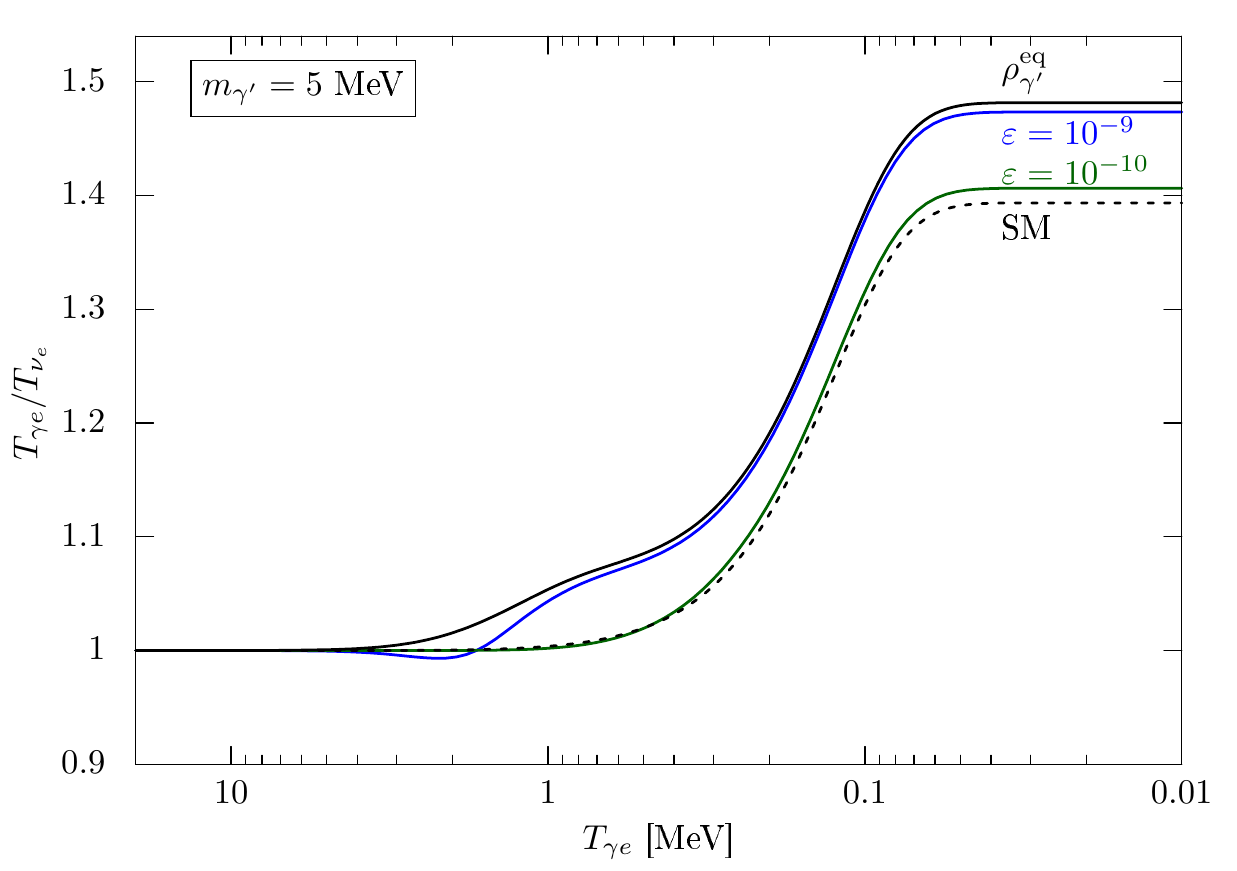}}
\caption{The time evolution of (a): the energy density of the dark photon $\rho_{\gamma'}$ and (b): the ratio of photon and neutrino temperatures for the freeze-in scenario.
Here we adopt $m_{\gamma'} = 5$ MeV and $\varepsilon=10^{-8}$ (red), $10^{-9}$ (blue) and $10^{-10}$ (green), respectively.
For comparison we show the energy density of the dark photon 
in the thermal equilibrium (black).
We solve the Boltzmann equation 
from temperature 300\,MeV which is well above 
the neutrino decoupling temperature.
The evolution of $T_{\gamma e}/T_{\nu_e}$ for $\varepsilon \gtrsim 10^{-8}$ is almost identical to the completely thermalized case.
}
\label{fig:evolution}
\end{figure}

\section{\texorpdfstring{$N_{\mathrm{eff}}$}{Lg} constraints}
\label{sec:constratins}

Here, we show the results of the Boltzmann equations.
For the freeze-in scenario with the empty dark photon in the early 
universe, the dark photon is mainly produced at the low temperature 
of $T_{\gamma e} \sim \max\left[m_{\gamma'},m_e\right]$.
In Fig.~\ref{fig:evolution}, we show the time evolution of the 
dark photon energy density which is obtained by solving the Boltzmann 
equation for $m_{\gamma'} = 5$\,MeV (left panel).
In the figure, we take $\varepsilon = 10^{-8}$, $10^{-9}$ and $10^{-10}$,
where the dark photon lifetime  is about $ 0.5\times (10^{-9}/\varepsilon)^2$\,sec.
The figure shows that the dark photon 
is thermalized for $\varepsilon = 10^{-8}$, while 
it deviates from the thermal equilibrium and exhibits the 
out-of-equilibrium decay for $\varepsilon = 10^{-10}$.
In each choice of $\varepsilon$, we find sizable amount of the dark photon energy density is released to the electron-photon
plasma below the neutrino decoupling temperature, $T_{\nu} \simeq 2$\,MeV~\cite{Escudero:2018mvt}.

\begin{figure}[tbp]
\centering{\includegraphics[width=0.5\textwidth]{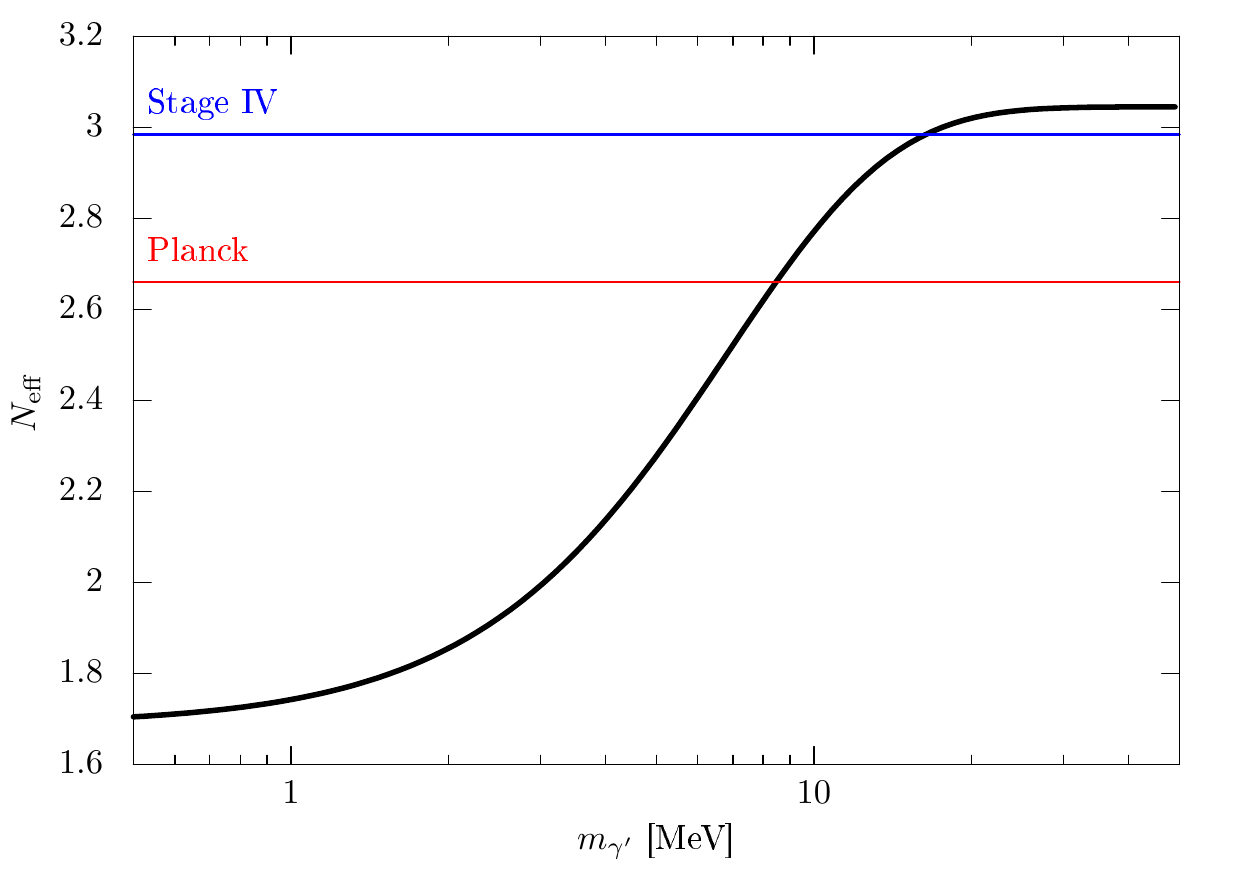}} 
\caption{
The $N_{\rm eff}$ as a function of the dark photon mass $m_{\gamma'}$.
Here we consider the case that the mixing $\varepsilon$ is large enough for the dark photon to be thermalized with the electron and photon.
In this case, the predicted $N_{\mathrm{eff}}$ does not 
depend on $\varepsilon$.
We shift the value of $N_{\mathrm{eff}}$ according to Eq.\,\eqref{eq:shift}.
}
\label{fig:thermal}
\end{figure}

In Fig.~\ref{fig:thermal}, we show the value of $N_{\mathrm{eff}}$
for the dark photon which completely freezes-in and is in the thermal equilibrium with the photon thermal bath.
Such a scenario is achieved for $\varepsilon \gg 10^{-8}$.
As the dark photon energy density follows the value in the 
thermal equilibrium (i.e. the black line in Fig.~\ref{fig:evolution}),
the predicted $N_{\mathrm{eff}}$ does not depend on $\varepsilon$.
The red line in the figure shows the lower limit 
of the present Planck constraint, $N_{\mathrm{eff}}=2.99^{+0.34}_{-0.33}$ at the 95\%CL~\cite{Aghanim:2018eyx}. 
The blue line shows the prospected sensitivity  
at the $2\sigma$ of the stage-I\!V CMB,
$\delta N_{\rm eff} = 0.06$ \cite{Abazajian:2016yjj}.
As a result, we find the 
dark photon mass $m_{\mathrm{\gamma'}} < 8.5$\,MeV has been excluded by the current Planck data for the completely freezed-in dark photon.
The stage-I\!V CMB observation will be also sensitive to the 
dark photon mass, $m_{\gamma'}\simeq 17$\,MeV.

\begin{figure}[htb]
	\centering
	\subcaptionbox{\label{fig:freeze-in_result} Freeze-in }{\includegraphics[width=0.47\textwidth]{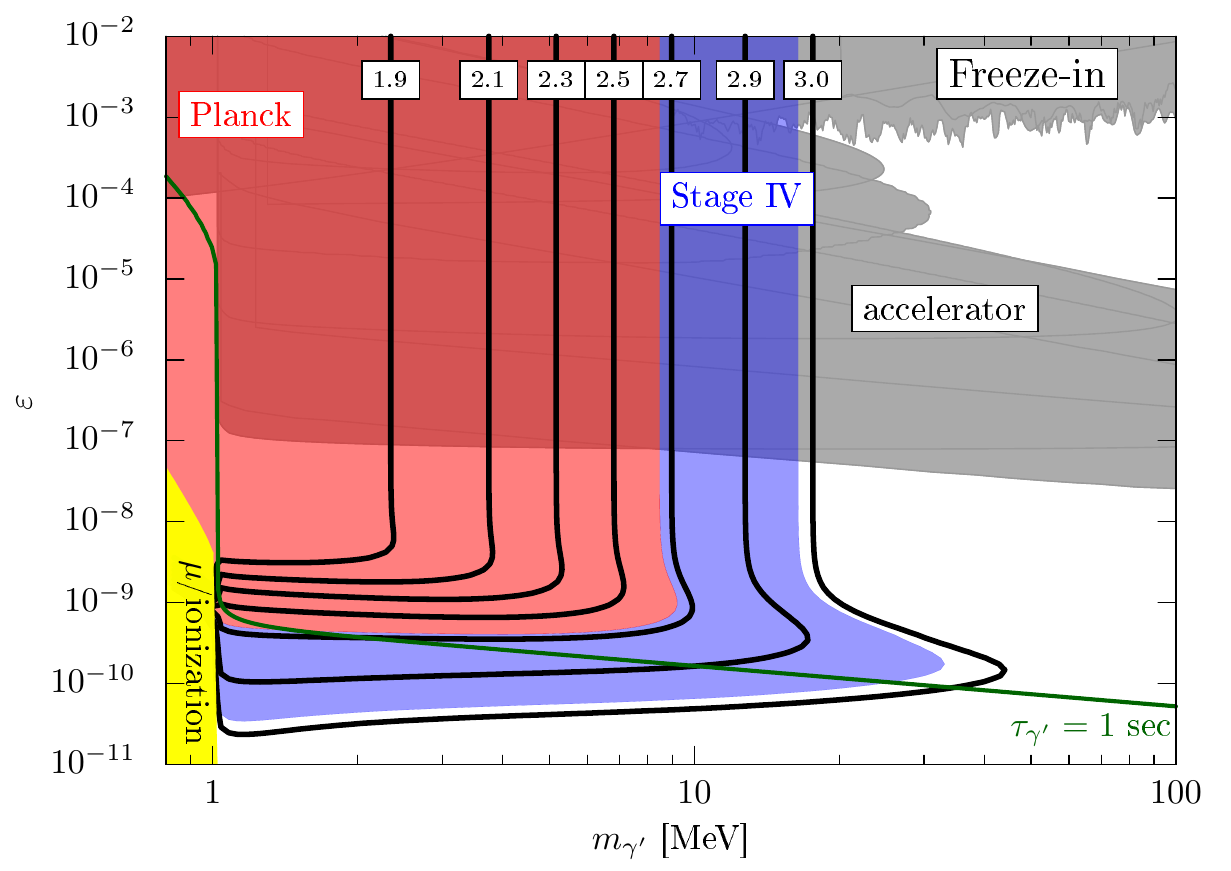}}
	\subcaptionbox{\label{fig:decoupling_result} Decoupling temperature $T_{D}=1$ TeV}{\includegraphics[width=0.47\textwidth]{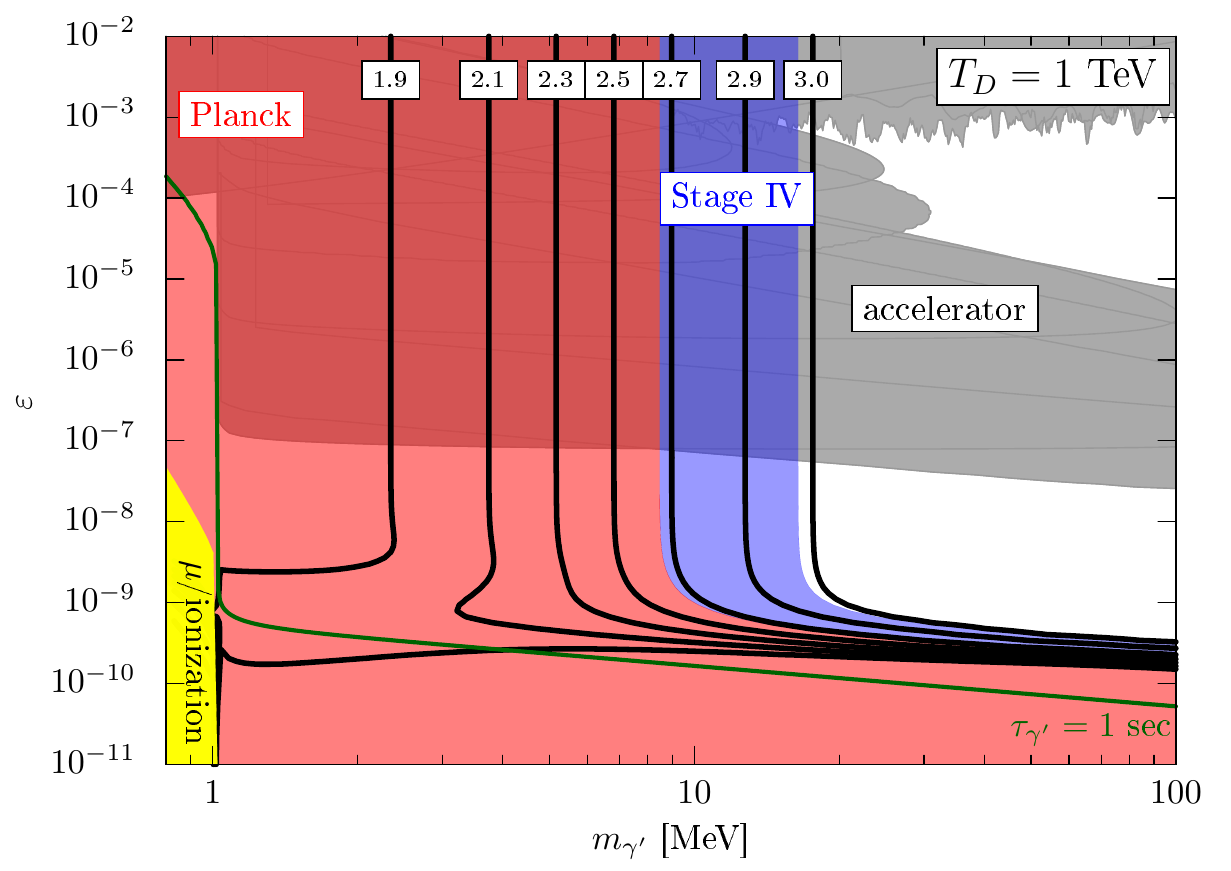}}
\caption{The contour plots of $N_{\rm eff}$ on the $m_{\gamma'}$-$\varepsilon$ plane.
The left panel corresponds to the 
freeze-in scenario, while the 
right panel to the early decoupled scenario.
The red region shows the present Planck constraint $N_{\rm eff} = 2.99^{+0.34}_{-0.33}~(95\%)$ \cite{Aghanim:2018eyx}.
The blue region shows that the sensitivity of the stage I\!V CMB experiment $\delta N_{\rm eff} = 0.06$ \cite{Abazajian:2016yjj}.
The yellow shaded region is excluded by the constraint on the CMB distortion ($\mu$ distortion)~\cite{Ellis:1990nb}
and the effects on the reionization history~\cite{Zhang_2007}.
The green line corresponds to the parameters where the lifetime of the dark photon
is $\tau_{\gamma'}=1$\,sec.
The gray shaded region is the compilation
of the constraints from the beam dump and the collider experiments in Ref.~\cite{Bauer_2018}, NA64 \cite{Banerjee:2019hmi}
and the electron $g-2$ constraint \cite{Hanneke:2008tm,Aoyama:2017uqe,Davoudiasl:2012ig,Endo:2012hp}.
}
\label{fig:result}
\end{figure}

In Fig.~\ref{fig:result}, we show the contour plot of $N_{\mathrm{eff}}$
on the $(m_{\gamma'}, \varepsilon)$ plane (left panel).
As we have mentioned above, the dark photon freezes-in completely 
for $\varepsilon \gg 10^{-8}$, and hence, the predicted 
$N_{\mathrm{eff}}$ does not depend on $\varepsilon$.
For a smaller $\varepsilon$, on the other hand, the dark photon 
is not completely freezed-in, and the dark photon effect 
on $N_{\mathrm{eff}}$ becomes small for $\varepsilon \ll 10^{-9}$.

In the figure, we shade 
the present Planck constraint (95\%CL) by red, 
while the stage-I\!V CMB sensitivity (2$\sigma$) is shaded by blue.
The figure shows the freeze-in dark photon is excluded for $\varepsilon \gtrsim 10^{-10}$ and $m_\gamma \lesssim 8.5$\,MeV
by the latest Planck constraint.
For $m_{\gamma'} < 2m_e$, the decay temperature of the dark photon becomes very low.
In the yellow shaded region, the decay temperature is lower than the $\mu$-distortion.
This region has been excluded by the 
constraint on the $\mu$ distortion~\cite{Ellis:1990nb} 
and
the effects on reionization history~\cite{Zhang_2007}.
We also show the robust constraints on the dark photon 
parameters based on the accelerator experiments,
SLAC E137, SLAC E141~\cite{Riordan:1987aw, Bjorken:1988as, Bjorken:2009mm, Andreas:2012mt},
Fermilab E774~\cite{Bross:1989mp},
Orsay~\cite{Davier:1989wz},
BaBar~\cite{Aubert:2009cp,Lees:2014xha},
A1~\cite{Merkel:2011ze},
KLOE~\cite{Archilli:2011zc, Babusci:2012cr, Anastasi:2016ktq, Anastasi:2015qla},
NA48/2~\cite{Batley:2015lha},
APEX~\cite{Abrahamyan:2011gv},
U70/Nu-Cal~\cite{Blumlein:2011mv, Blumlein:2013cua},
CHARM~\cite{Bergsma:1985is},
LSND~\cite{Athanassopoulos:1997er},
which are compiled in Ref.~\cite{Bauer_2018}, NA64 \cite{Banerjee:2019hmi}%
and the electron $g-2$ constraint \cite{Hanneke:2008tm,Aoyama:2017uqe,Davoudiasl:2012ig,Endo:2012hp}.

It should be noted that the dark photon effect on $N_{\mathrm{eff}}$
is enhanced at around $\varepsilon= \order{10^{-10}}$ where 
the dark photon decays in an out-of-equilibrium way.
As the dark photon is a massive particle, the relative energy density of 
the dark photon is enhanced by the cosmic expansion by the time of the decay, which enlarges the dark photon effect on $N_{\mathrm{eff}}$.
As a result, we find that the stage-I\!V CMB is sensitive to 
the dark photon mass of $m_{\gamma'}\simeq 30$\,MeV for $\varepsilon \simeq 10^{-10}$.

Light particles produced in a supernova explosion can alter the neutrino burst spectrum.
Thus, dark photons with $m_{\gamma'}\lesssim 100\, \mathrm{MeV}$ are constrained from the observation of neutrino burst of SN1987A~\cite{Bjorken:2009mm, Dent:2012mx, Kazanas:2014mca, Rrapaj:2015wgs, Chang:2016ntp}.
Recently, however, it is pointed out that there are uncertainties in a model of the neutrino burst and there is a possibility that the constraints from SN1987A are discarded~\cite{Bar:2019ifz}.
In view of such astrophysical uncertainties, we do not show the supernova constraints.

Next, we consider the dark sector freeze-out scenario where the dark and the SM 
sector thermal bath are in the equilibrium in the very early 
universe and then decouple at a certain low temperature.
In the right panel 
of Fig.~\ref{fig:result},
we show $N_{\mathrm{eff}}$ for the 
dark sector freeze-out scenario of $T_{D}= 1$\,TeV.
In this case, the initial condition of $f_{\gamma'}$
is given by Eq.\,\eqref{eq:decoupled} at $T_{\gamma e} \gg 
\max[m_e,m_{\gamma'},T_{FI}]$.
Due to the preexisting dark photon 
abundance well before the freeze-in production,
the parameter region with $\tau_{\gamma'} = \order{1}$\,sec has been excluded by 
the Planck constraint.
If the dark sector decoupling temperature gets lower, the constraint gets stronger.
The reason follows: in the case that the
dark sector freeze out occurs at lower temperature, $g^*_s(T_D)$ gets smaller, which makes the energy density of the dark photon larger.
Then the resultant energy injection into the electron-photon plasma gets larger and the constraint becomes severer.
For the parameter region with $\tau_{\gamma'} \ll \order{1}$\,sec,
the constraint is identical with that 
in the freeze-in scenario.

For the freeze-out scenario,
the $N_{\mathrm{eff}}$ constraint is valid
as long as $m_{\gamma'} \ll T_D$.
For $m_{\gamma'} \gg T_D$, the abundance produced from the dark sector thermal bath is suppressed by the Boltzmann factor, thus no constraint is imposed.
For $m_{\gamma'} \lesssim T_D$, all the parameter region
with a tiny $\varepsilon$ where $\tau_{\gamma'} \gg \order{1}\,$sec is excluded.

Finally, let us comment on the $X$(16.7 MeV) boson, which is reported in 
the
${}^8{\rm Be^*}$ nucleus decay, i.e., the so-called Berillium anomaly~\cite{Krasznahorkay:2015iga,Krasznahorkay:2018snd}.
To explain this anomaly, the hidden vector boson is required to have a
sizable coupling to the electrons~\cite{Feng:2016ysn,Feng:2016jff} and thus we can apply the current $N_{\rm 
eff}$ constraint.
For $X$(16.7 MeV), the predicted $\Delta N_{\rm eff} \simeq 0.6$ and is 
still consistent with the current Planck data but can be probed with the future CMB stage-I\!V experiment.

\section{Discussions}
In this paper, we studied the $N_\mathrm{eff}$ constraint on the dark photon in detail.  As the MeV dark photon coupling to the neutrinos is suppressed, the dark photon heats up only the electron-photon plasma and reduces $N_{\mathrm{eff}}$, if it decays or annihilates after the neutrino decoupling.  For the dark photon mass above the electron-positron threshold, which is the main interest in this paper, we solve the Boltzmann equations of the dark photon with the Pauli-blocking and the Bose-enhancement fully included. We estimated the effects of the energy injection after the neutrino decoupling caused by the decay or annihilation of the MeV-scale dark photon.  As a result, we found that this effect leads to the decrease of the $N_{\mathrm{eff}}$, and the CMB stage-I\!V experiments can test a wide range of the MeV-scale dark photon.

Although we considered the Boltzmann equations which include the Pauli-blocking and the Bose-enhancement for $m_{\gamma'} > 2 m_e$, we have not taken into account the thermal effects on the kinetic mixing~\cite{Redondo:2008ec,Fradette:2014sza}. Such effects could slightly enhance the production rate of the dark photon. We also adopt an approximated treatment of the thermal mass effects to take care of the IR singularity for the scattering processes for $m_{\gamma'} > 2 m_e$. The analysis with the full Boltzmann equations of the dark photon momentum distribution with those effects requires consistent treatment of the thermal effects including the higher-order corrections, 
though we expect such effects are insignificant since the dark photon production is dominated by 
the decay and the inverse decay process
for $m_{\gamma'}>2m_e$.

\section*{Acknowledgments}

The authors thank K. Inomata for a useful comment on the $\mu$-distortion constraints.
This work is supported by Grant-in-Aid for Scientific Research from the Ministry of Education, Culture, Sports, Science, and Technology (MEXT), Japan, 17H02878 (M.I. and S.S.), 15H05889, 16H03991  17H02878, 18H05542 (M.I.) 18K13535, 19H04609 (S.S.), and by World Premier International Research Center Initiative (WPI), MEXT, Japan. 
This work is also supported by the Advanced Leading Graduate Course for Photon Science (S.K.) and International Graduate Program for Excellence in Earth-Space Science (Y.N.).

\appendix
\section{Dark Photon Decay Collision Term}
\label{sec:DP decay}
Here, we show the explicit calculation of the collision term of dark photon decay Eqs.~\eqref{eq:boltzmann},\eqref{eq:DecayInverseDecay},\eqref{eq:phi}.

First, the collision term is defined as
\begin{align}
    G_{ \gamma' \leftrightarrow e}(p_{\gamma'},T) &= \frac{1}{2E_{\gamma'}}\int \frac{g_{e^-}d^3p_{e^-}}{(2\pi)^32E_{e^-}}\frac{g_{e^+}d^3p_{e^+}}{(2\pi)^32E_{e^+}}(2\pi)^4\delta^4(p_{\gamma'}-p_{e^-}-p_{e^+})\abs{\bar{\mathcal{M}}}^2\notag \\
    &\hspace{2cm}\times\qty[f_{\gamma'}(1-f_{e^-}(T))(1-f_{e^+}(T))-(1+f_{\gamma'})f_{e^-}(T)f_{e^+}(T)]\notag \\
    &= \frac{\abs{\bar{\mathcal{M}}}^2}{2E_{\gamma'}}[f_{\gamma'}F_1(E_{\gamma'},T)-F_2(E_{\gamma'},T)].
\end{align}
Here we defined $F_1, F_2$ as
\begin{align}
    F_1(E_{\gamma'},T) &= \int \frac{g_{e^-}d^3p_{e^-}}{(2\pi)^32E_{e^-}}\frac{g_{e^+}d^3p_{e^+}}{(2\pi)^32E_{e^+}}(2\pi)^4\delta^4(p_{\gamma'}-p_{e^-}-p_{e^+})(1-f_{e^-}(T)-f_{e^+}(T))\\
    F_2(E_{\gamma'},T) &= \int \frac{g_{e^-}d^3p_{e^-}}{(2\pi)^32E_{e^-}}\frac{g_{e^+}d^3p_{e^+}}{(2\pi)^32E_{e^+}}(2\pi)^4\delta^4(p_{\gamma'}-p_{e^-}-p_{e^+})f_{e^-}(T)f_{e^+}(T),
\end{align}
where $g_{e^\pm}=2$ is the spin degree of freedom of $e^\pm$ and 
\begin{align}
    \abs{\bar{\mathcal{M}}}^2 = \frac{4\pi}{3}\alpha\varepsilon^2m_{\gamma'}^2\qty(1+\frac{2m_e^2}{m_{\gamma'}^2})=\frac{4\pi m_{\gamma'}\Gamma_{\gamma'}}{\sqrt{1-4m_e^2/m_{\gamma'}^2}}=\frac{2\pi m_{\gamma'}^2\Gamma_{\gamma'}}{p^0_e}
\end{align}
is the spin-averaged amplitude squared.
By doing the the integral of the $\delta$-function, $F_1$ is reduced to
\begin{align}
    F_1(E_{\gamma'},T) &= \frac{1}{2\pi p_{\gamma'}}\int_{E_e^-}^{E_e^+}dE_e(1-f_{e^-}(E_e,T)-f_{e^+}(E_{\gamma'}-E_e,T))\\
    E_e^\pm &= \frac{E_{\gamma'}}{2}\qty(1\pm \sqrt{1-\frac{4m_e^2}{m_{\gamma'}^2}}\sqrt{1-\frac{m_e^2}{m_{\gamma'}^2}})=\frac{1}{m_{\gamma'}}(E_{\gamma'}E^0_e\pm p_{\gamma'}p^0_e).\notag
\end{align}
In the text, we assumed that electrons obey the FD distribution, thus we can explicitly do the integral
\begin{align}
    F_1(E_{\gamma'},T) &= \frac{1}{2\pi p_{\gamma'}}\int_{E_e^-}^{E_e^+}dE_e\qty(1-\frac{1}{e^{E_e/T}+1}-\frac{1}{e^{(E_{\gamma'}-E_e)/T}+1})\notag \\
    &=\frac{T}{2\pi p_{\gamma'}}\eval{\ln\qty(\frac{e^{E_e/T}+1}{e^{E_{\gamma'}/T}+e^{E_e/T}})}_{E_e=E_e^-}^{E_e=E_e^+}\notag \\
    &= \frac{T}{2\pi p_{\gamma'}}\times \frac{2p_{\gamma'}p^0_e}{m_{\gamma'}T}(1+\varphi(T,p_{\gamma'}))\notag \\
    &=\frac{p_e^0}{\pi m_{\gamma'}}(1+\varphi(T,p_{\gamma'})),
\end{align}
where $\varphi$ is defined in Eq.~\eqref{eq:phi}.
In the same manner, we can show that
\begin{align}
    F_2(E_{\gamma'},T) = F_1(E_{\gamma'},T)f_{\gamma'}^{\mathrm{eq}}(T).
\end{align}
Thus we can derive that
\begin{align}
    G_{ \gamma' \leftrightarrow e}(p_{\gamma'},T) &= \frac{\abs{\bar{\mathcal{M}}}^2}{2E_{\gamma'}}F_1(E_{\gamma'},T)(f_{\gamma'}-f_{\gamma'}^{\mathrm{eq}}(T))\notag \\
    &=\frac{m_{\gamma'}\Gamma_{\gamma'}(1+\varphi(T,p_{\gamma'}))}{E_{\gamma'}}(f_{\gamma'}-f_{\gamma'}^{\mathrm{eq}}(T)).
\end{align}

\section{Neutrino-Electron Collision Terms}
\label{sec:neutrino collision}
Here, we present the definition of the collision terms and the explicit forms of
the electron-neutrino and the neutrino-neutrino collision terms.

First, the general form of the Boltzmann equation for the process $\psi + i \leftrightarrow f$ is given by
\begin{align}
    \pdv{f_\psi}{t}-H p \pdv{f_\psi}{p}
    &=\mathcal{C}_{\psi+X \leftrightarrow Y}[ f_\psi ]\ ,\\
    \mathcal{C}_{\psi+X \leftrightarrow Y}[f_\psi] &= -\frac{1}{2E_\psi}\int d \Pi_X d \Pi_Y (2\pi)^4\delta \qty(p_\psi + \sum_i p_{X_i} - \sum_j p_{Y_j}) S\abs{\bar{\mathcal{M}}}^2\notag \\
	&\hspace{1cm} \times \qty[f_\psi \prod_i f_{X_i} \prod_j (1\pm f_{Y_j}) - (1\pm f_\psi)\prod_i (1\pm f_{X_i})\prod_j f_{Y_j}]\ .   \\
	d \Pi_i d \Pi_f &= \prod_i\frac{g_{X_i} d^3 p_{X_i}}{(2\pi)^3 2E_{X_i}} \prod_j\frac{g_{Y_j} d^3 p_{Y_j}}{(2\pi)^3 2E_{Y_j}} \ ,
\end{align}
where $X$ and $Y$ can be multi-particle states and $g$ is a spin degrees of freedom of each particle. 
$\abs{\bar{\mathcal{M}}}^2$ is a amplitude squared averaged over spin degrees of freedom of all particles in $\psi +X$ and $Y$.
The factor $S$ is a symmetrization factor which gives $1/2!$ for each pair of identical particles in $X$ and $Y$.%
\footnote{Note that $S$ gives a factor $1/2!$ for each pair of identical particles in $X$, not in $\psi + X$.
If $X$ is a one-particle state, $X = \psi$, $S$ simply gives factor $1$.
Also, note that if there is a set $n$ identical particles, $S$ gives a factor $1/n!$ for each set.}

In the text, we define
\begin{align}
    G_{\gamma' \leftrightarrow e} &= \mathcal{C}_{\gamma' \leftrightarrow e^+e^-}\ , \\
    C_{e \leftrightarrow \nu_i} &= \int \frac{g_e d^3 p_e}{(2\pi)^3}
                                    \qty(\mathcal{C}_{e^+e^-\leftrightarrow \nu_i \bar{\nu}_i} +\mathcal{C}_{e^\pm\nu_i\leftrightarrow e^\pm\nu_i} +\mathcal{C}_{e^\pm\bar{\nu}_i\leftrightarrow e^\pm\bar{\nu}_i})\ , \\
    C_{\nu_i \leftrightarrow \nu_j} &= \int \frac{g_e d^3 p_e}{(2\pi)^3}
                                    \qty(\mathcal{C}_{\nu_i\nu_j\leftrightarrow \nu_i\nu_j}+\mathcal{C}_{\nu_i\bar{\nu}_i\leftrightarrow \nu_j\bar{\nu}_j})\ .\\
\end{align}
According to Ref~\cite{Escudero:2018mvt,Ichikawa:2005vw,Kawasaki:2000en}, the $e\mbox{--}\nu$ collision terms are written in the form of
\begin{align}
    C_{e \leftrightarrow \nu_e }(T_{\gamma e}, T_{\nu_{e}}) 
    &=-\frac{G_{\mathrm{F}}^2}{\mathrm{\pi}^5}
    (1 + 4 s_{\mathrm{W}}^2 + 8 s_{\mathrm{W}}^4)F(T_{\gamma e}, T_{\nu_e})\ , \\
    C_{e \leftrightarrow \nu_{\mu,\tau}}(T_{\gamma e}, T_{\nu_\mu})
    &=-\frac{G_{\mathrm{F}}^2}{\mathrm{\pi}^5}
    (1 - 4 s_{\mathrm{W}}^2 + 8 s_{\mathrm{W}}^4)F(T_{\gamma e}, T_{\nu_{\mu}})\ ,\\
    F(T_1,T_2) &= 32(T_1^9 - T_2^9) + 56 T_1^4 T_2^4(T_1 - T_2)\ ,
\end{align}
where $G_F$ is the Fermi constant.
Using the same $F$ function, we can write the $\nu_e\mbox{--}\nu_{\mu,\tau}$ term as
\begin{align}
    C_{\nu_e \leftrightarrow \nu_{\mu, \tau} }(T_{\nu_e}, T_{\nu_\mu})
    &=-\frac{G_{\mathrm{F}}^2}{\mathrm{\pi}^5}
    F(T_{\nu_e}, T_{\nu_\mu})\ .
\end{align}

\section{Dark Photon Production via Scattering}
\label{sec:scattering}
In this appendix, we summarize the collision terms of the
dark photon through the electron scattering, 
\begin{align}
    \mathcal{C}_{sc}[f_{\gamma'}] =&-\frac{1}{2E_{\gamma'}}\int 
    \frac{2d^3\ps_3}{(2\pi)^32E_3}
    \frac{2d^3\ps_2}{(2\pi)^32E_2}
    \frac{2d^3\ps_{1}}{(2\pi)^32E_{1}}(2\pi)^4 \delta^4(p_{3}+p_{\gamma'}-p_{1}-p_{2})
   \times \cr
    &\left[
    |\bar{\mathcal{M}}_{e^+e^-\leftrightarrow\gamma\gamma'}|^2 
    (f_{\gamma}f_{\gamma'}(1-f_e)(1-f_{\bar{e}})- f_e f_{\bar{e}}(1+f_{\gamma})(1+f_{\gamma'}))\right. \cr
& + 
|\bar{\mathcal{M}}_{e^-\gamma\leftrightarrow e^-\gamma'}|^2
    (f'_{e}f_{\gamma'}(1-f_e)(1+f_{\gamma})- f_e f_{\gamma}(1-f'_e)(1+f_{\gamma'}))\cr
& +
    |\bar{\mathcal{M}}_{e^+\gamma\leftrightarrow e^+\gamma'}|^2
    (f'_{\bar{e}}f_{\gamma'}(1-f_{\bar{e}}) (1+f_{\gamma})
    - f_{\bar{e}} f_{\gamma}(1-f'_{\bar{e}})(1+f_{\gamma'}))
    \left.
    \right]\ . 
\end{align}
Here, $|\bar{\mathcal{M}}|^2$ denotes the squared amplitude with all the 
spins averaged.
We put $^\prime$ on the distribution for a later use, although 
$f'=f$.

In the following, we use the Maxwell-Boltzmann (MB) approximation in the following way.
For $ee\leftrightarrow\gamma\gamma'$, we approximate
\begin{align}
    & (f_{\gamma}f_{\gamma'}(1-f_e)(1-f_{\bar{e}})- f_e f_{\bar{e}}(1+f_{\gamma})(1+f_{\gamma'})) \cr
   &  \to 
   (f_{\gamma}f_{\gamma'}- f_e f_{\bar{e}}(1+f_{\gamma})(1+f_{\gamma'})) \ ,
   \label{eq:feeggp}
\end{align}
where $f_e$ and $f_{\bar{e}}$ are the MB distribution
while $f_\gamma$ is taken to be the BE distribution.
For $e\gamma\leftrightarrow e\gamma'$, on the other hand, we take
\begin{align}
    & (f'_{{e}}f_{\gamma'}(1-f_{{e}}) (1+f_{\gamma})
    - f_{{e}} f_{\gamma}(1-f'_{{e}})(1+f_{\gamma'}))\cr
   & \to  (f'_{{e}}f_{\gamma'}
    - f_{{e}} f_{\gamma}(1-f'_{{e}})(1+f_{\gamma'}))\ .
\end{align}
where $f_{e}$ and $f_\gamma$ are the MB distribution, while $f'_e$ is the FD distribution.
With these approximations, the distribution of $\gamma'$ converges to the BE distribution in the
equilibrium limit.

Therefore, we obtain
the collision term as,
\begin{align}
    \mathcal{C}_{sc}[f_{\gamma'}] = -
\left(
\tilde G_{e^+e^-\leftrightarrow\gamma\gamma'}
+
\tilde G_{e^-\gamma\leftrightarrow e^-\gamma'}
+
\tilde G_{e^+\gamma\leftrightarrow e^+\gamma'}
\right)\times
\left(f_{\gamma'}(E_{\gamma'})- \frac{1}{e^{E_{\gamma'}/T}-1}\right)\ ,
\end{align}
where
\begin{align}
 \tilde G_{e^+e^-\leftrightarrow\gamma\gamma'}  =  \frac{1}{2E_{\gamma'}f^{{BE}}_{\gamma'}(E_{\gamma'})}\int &
    \frac{2d^3\ps_3}{(2\pi)^32E_3}
    \frac{2d^3\ps_2}{(2\pi)^32E_2}
    \frac{2d^3\ps_{1}}{(2\pi)^32E_{1}}(2\pi)^4 \delta^4(p_{3}+p_{\gamma'}-p_{1}-p_{2}) \cr
    &\times|\bar{\mathcal{M}}_{e^+e^-\leftrightarrow\gamma\gamma'}|^2
    f_e f_{\bar{e}}(1+f_{\gamma})\ ,
    \cr
    \tilde G_{e^-\gamma\leftrightarrow e^-\gamma'}  =  \frac{1}{2E_{\gamma'}f^{{BE}}_{\gamma'}(E_{\gamma'})}\int &
    \frac{2d^3\ps_3}{(2\pi)^32E_3}
    \frac{2d^3\ps_2}{(2\pi)^32E_2}
    \frac{2d^3\ps_{1}}{(2\pi)^32E_{1}}(2\pi)^4 \delta^4(p_{3}+p_{\gamma'}-p_{1}-p_{2})
  \cr
    &\times|\bar{\mathcal{M}}_{e^-\gamma\leftrightarrow e^-\gamma'}|^2
    f_e f_{\gamma}(1-f'_{e})\ .
\end{align}

Following Ref.~\cite{Bae:2017dpt}, the above integration can be further reduced to
\begin{align}
 \tilde G_{e^+e^-\leftrightarrow\gamma\gamma'}  
    = \frac{1}{512\pi^3} \frac{T e^{-E_{\gamma'}/T}}{|\ps_{\gamma'}|E_{\gamma'}f_{\gamma'}(E_{\gamma'})}\int & ds  \frac{1}{\sqrt{s} |\ps^{\mathrm{cms}}_{\gamma\gamma'}|}
 \log\left[\frac{1-e^{-E_\gamma^+/T}}{1-e^{-E_\gamma^-/T}}\right]
    \int dt |\bar{\mathcal{M}}_{e^+e^-\leftrightarrow\gamma\gamma'}|^2\times 8\ ,
    \cr
    \tilde G_{e^-\gamma\leftrightarrow e^-\gamma'}  
  =  \frac{1}{512\pi^3} \frac{T e^{-E_{\gamma'}/T}}{|\ps_{\gamma'}|E_{\gamma'}f_{\gamma'}(E_{\gamma'})}\int & ds  \frac{1}{\sqrt{s} |\ps^{\mathrm{cms}}_{e\gamma'}|}
    \log\left[\frac{1+e^{-E_e^-/T}}{1+e^{-E_e^+/T}}\right]
 \int dt |\bar{\mathcal{M}}_{e^-\gamma\leftrightarrow e^-\gamma'}|^2
    \times 8\ .
\end{align}
Here, 
\begin{align}
   & E_\gamma^\pm= \frac{s-m_{\gamma'}^2}{2(E_{\gamma'}\mp|\ps_{\gamma'}|)}\ ,\cr
   & E_e^\pm  =\frac{{E}_{\gamma '} \left(s-m_e^2-m_{\gamma '}^2\right)\pm|\ps_{\gamma'}|\sqrt{\left(s-(m_e + m_{\gamma'})^2\right)\left(s-(m_e - m_{\gamma'})^2\right)}}{2
m_{\gamma '}^2}\ , \cr
&
  |\ps^{\mathrm{cms}}_{\gamma\gamma'}| = \frac{\sqrt{(s-(m_{\gamma'})^2)(s-(m_{\gamma'})^2)}}{2\sqrt{s}} \ , \cr
&   |\ps^{\mathrm{cms}}_{e\gamma'}| = \frac{\sqrt{(s-(m_e+m_{\gamma'})^2)(s-(m_e-m_{\gamma'})^2)}}{2\sqrt{s}} \ , 
\end{align}
respectively.

The $t$-integration of the spin averaged squared matrices are given by
\begin{align}
\label{eq:eegammagammap}
\int_{t_\mathrm{min}}^{t_\mathrm{max}}dt&|\bar{\mathcal M}|_{e^+e^-\to\gamma\gamma'}^2=
 \frac{\varepsilon^2 g^4}{3\sqrt{s}(s-m_{\gamma '}^2)}  \left[
 -2 \sqrt{s-4 m_e^2} \left(4 s m_e^2+m_{\gamma '}^4+s^2\right)
   \right. \cr
   &
   \left.
   + 2 \sqrt{s} \log \left(\frac{\sqrt{s \left(s-4 m_e^2\right)}-2 m_e^2+s}{2
   m_e^2}\right) \left(4 m_e^2 \left(s-m_{\gamma '}^2\right)-8 m_e^4+m_{\gamma
   '}^4+s^2\right)
\right]
  \ ,
\end{align}
and 
\begin{align}
    \int_{t_\mathrm{min}}^{t_\mathrm{max}}&dt|\bar{\mathcal M}|_{e^+\gamma\to e^+\gamma'}^2=
    \frac{\varepsilon^2g^4}{6s(s-m_e^2)}\cr
    &
  \times  \left[
  \beta  \left(-m_e^4 \left(m_{\gamma '}^2+s\right)+s m_e^2 \left(2 m_{\gamma
   '}^2+15 s\right)+m_e^6+s^2 \left(7 m_{\gamma '}^2+s\right)\right)
   \right.\cr
   & \left.
    +
   4s\left(m_e^2 \left(2 m_{\gamma '}^2-6 s\right)-3 m_e^4+2 m_{\gamma '}^4-2 s
   m_{\gamma '}^2+s^2\right) \log \left(\frac{m_e^2-m_{\gamma '}^2+\beta  s+s}{2
   \sqrt{s} m_e}\right)
   \right]\ ,
\end{align}
respectively.
Here, $\beta$ is defined by
\begin{align}
\beta s = \sqrt{(s-(m_e+m_{\gamma'})^2)(s-(m_e-m_{\gamma'})^2)}\ .
\end{align}
These results are consistent with those of Ref.~\cite{Redondo:2008ec}.

It should be noted that the factor $(s-m_{\gamma'}^2)^{-1}$ appearing 
in Eq.\,\eqref{eq:eegammagammap} causes 
a linear IR divergence for $m_{\gamma'}> 2m_e$.%
\footnote{If we use MB distributions for $f_\gamma$ in Eq.\,\eqref{eq:feeggp},
the IR divergence is logarithmic.} 
To regulate the IR divergence,
we adopt the prescription in Ref.~\cite{Redondo:2008ec}
where this factor is replaced by 
\begin{align}
    (s-m_{\gamma'}^2)^{-1} \to \left(s-m_{\gamma'}^2
    +2 m_{\gamma'}\delta m_{\gamma}(T)\right)^{-1}\ .
\end{align}
Here, $\delta m_{\gamma}^2(T)$ is the thermal photon mass given in Eq.\,\eqref{eq:thermal photon mass}.

Finally, let us comment on our treatment on the decay 
of dark photon for $m_{\gamma'} < 2m_e$. 
As our main interest in the present paper is on $m_{\gamma'}>2m_e$,
we adopt an approximate treatment of the three-body for $m_{\gamma'} < 2 m_e$,
\begin{align}
    \mathcal{C}_{\gamma'\leftrightarrow 3\gamma}[f_{\gamma'}] = - \frac{m_{\gamma'}}{E_{\gamma'}}\Gamma_{\gamma'\to 3\gamma}\times 
    \left(
    f_{\gamma'}(E_{\gamma'}) - \frac{1}{e^{E_{\gamma'}/T}-1}
    \right)\ ,
\end{align}
where $\Gamma_{\gamma'\to 3\gamma}$ is in given Eq.\,\eqref{eq:3gamma}.

\bibliographystyle{apsrev4-1}
\bibliography{ref}

\end{document}